\documentclass[%
 reprint,
nofootinbib,
 amsmath,amssymb,
 aps
]{revtex4-1}
\pdfoutput=1
\usepackage{graphicx}% Include figure files
\usepackage[utf8]{inputenc}
\usepackage{flushend}
\usepackage{dcolumn}% Align table columns on decimal point
\usepackage{bm}% bold math
\usepackage{balance}
\usepackage[normalem]{ulem}

\usepackage[colorlinks = true,
            linkcolor = blue,
            urlcolor  = blue,
            citecolor =green,
            anchorcolor = blue]{hyperref}
\usepackage{verbatim}
\usepackage{color,ulem}
\usepackage[english]{babel}

\usepackage[utf8]{inputenc}

\newcommand{\beq}{\begin{eqnarray}}
\newcommand{\eeq}{\end{eqnarray}}
\usepackage{amsmath}
\usepackage{tikz}
\usetikzlibrary{decorations.pathmorphing}
\usetikzlibrary{shapes.misc}
\tikzset{cross/.style={cross out, draw=black, minimum size=8*(#1-\pgflinewidth), inner sep=0pt, outer sep=0pt},
cross/.default={1pt}}
\usetikzlibrary{patterns,math}
\begin{document}

\title{\Large A new paradigm for the low-$T$ glassy-like thermal properties of solids}

%\title{\Large Universal origin of low-$T$ specific heat anomalies in crystals, quasicrystals, and glasses}

\author{Matteo Baggioli}%
 \email{matteo.baggioli@uam.es}
\affiliation{Instituto de Fisica Teorica UAM/CSIC, c/Nicolas Cabrera 13-15,
Universidad Autonoma de Madrid, Cantoblanco, 28049 Madrid, Spain.
}%

\author{Alessio Zaccone}%
 \email{alessio.zaccone@unimi.it }
\affiliation{
Department of Physics "A. Pontremoli", University of Milan, via Celoria 16, 20133 Milan, Italy.\\
Department of Chemical Engineering and Biotechnology,
University of Cambridge, Philippa Fawcett Drive, CB30AS Cambridge, U.K.\\
Cavendish Laboratory, University of Cambridge, JJ Thomson
Avenue, CB30HE Cambridge, U.K.
}

\begin{abstract}
Glasses and disordered materials are known to display anomalous features in the density of states, in the specific heat and in thermal transport. Nevertheless, in recent years, the question whether these properties are really anomalous (and peculiar of disordered systems) or rather more universal than previously thought, has emerged. New experimental and theoretical observations have questioned the origin of the boson peak and the linear in $T$ specific heat exclusively from disorder and TLS. The same properties have been indeed observed in ordered or minimally disordered compounds and in incommensurate structures for which the standard explanations are not applicable. Using the formal analogy between phason modes (e.g. in quasicrystals and incommensurate lattices) and diffusons, and between amplitude modes and optical phonons, we suggest the existence of a more universal physics behind these properties. In particular, we strengthen the idea that linear in $T$ specific heat is linked to low energy diffusive modes and that a BP excess can be simply induced by gapped optical-like modes.

\end{abstract}

\maketitle
%\section*{Introduction}
\textbf{\textit{Introduction}} -- Anomalous low-temperature thermal properties play a big role in the mysterious nature of the glassy state -- which is considered ''perhaps the deepest and most interesting unsolved problem in solid state theory''. On one side, glasses exhibit a linear in temperature contribution to the non-electronic specific heat \cite{PhysRevB.4.2029} at low $T$; from the other, together with various disordered materials, they display a characteristic excess in the vibrational density of states (VDOS) -- the boson peak (BP) --  which violates the standard result from Debye theory, i.e. $C(T)\sim T^3$ \cite{PhysRevB.4.2029,Shintani2008,PhysRevB.87.134203,PhysRevLett.96.045502}.\\

The most (and probably only) accepted theory to explain the linear in $T$ contribution relies on so-called \textit{two-level states} (TLS) \cite{Phillips1972,Anderson1972-ANDALT-2,1987RPPh...50.1657P}, while many theoretical explanations for the BP anomaly have been proposed \cite{Ruocco2007,Taraskin,PhysRevB.43.5039} over the past decades. Importantly, all these theoretical frameworks rely strongly on disorder in various forms and on the existence of new low energy degrees of freedom, the exact physical nature of which has remained somewhat mysterious \cite{doi:10.1021/jp402222g}.\\

In addition to this questionable confidence and theoretical confusion, the ``anomalous" nature of glassy features in solids is in deep crisis \cite{ramos2020universal}. This is because the same allegedly ``anomalous'' features have been systematically observed in non-disordered systems, atomic and molecular cryocrystals \cite{Krivchikov,PhysRevLett.119.215506,PhysRevB.99.024301} and in quasicrystals/incommensurate structures \cite{PhysRevLett.114.195502}.\\

Not only these features appear to be quite universal rather than anomalous, and not limited to glasses, but the chase for the universal physics or unifying principles behind these phenomena is completely open. In particular, neither disorder nor Two Level Systems (TLS) can be the ultimate causes of these phenomena since obviously no disorder and no TLS can be found in ordered crystals and in quasicrystals.\\

Recently, the idea that these (not anymore anomalous) features can actually be induced by different physical mechanisms has gained strength, supported by experimental observations \cite{Krivchikov,PhysRevLett.119.215506,PhysRevB.99.024301,PhysRevLett.114.195502} and first principles theoretical analyses. Robust theoretical frameworks, which fit harmoniously with several experimental data \cite{PhysRevResearch.2.013267} and observations \cite{Krivchikov,PhysRevLett.119.215506,PhysRevB.99.024301,PhysRevLett.114.195502}, have proven that a BP anomaly in crystals can be induced by overdamped low-energy modes and (Akhiezer-type) anharmonicity \cite{PhysRevLett.122.145501,casella2020physics} and also by the piling up of softly gapped optical-like modes \cite{Baggioli_2019,PhysRevB.100.220201}. Moreover, it has been shown \cite{PhysRevResearch.1.012010} that low-energy diffusive excitations produce a linear in $T$ contribution to the specific heat, in agreement with the results from random matrix theory \cite{PhysRevE.100.062131}. \\

In this Letter, we provide more evidence for the case by considering the low-temperature thermal properties of quasicrystals and incommensurate structures. Both anomalous features (BP and linear in T  specific heat) have been experimentally observed in quasicrystals \cite{PhysRevLett.114.195502}. Phenomenological models have been constructed and pinpoint as the origin of these anomalies the dynamics of the phason and amplitude modes \cite{PhysRevLett.93.245902,PhysRevB.70.212301}. Here, we show that everything can be explained in a universal, simple and elegant way using first principles ideas about the underlying symmetries.\\

The quasicrystal case provides another direct proof against the anomalous nature of glasses and substantial evidence towards the idea that disorder and TLS cannot be the universal causes behind these phenomena. The key of universality simply lies in the \textit{formal} nature (i.e. mathematical dispersion relations) of the low-energy excitations from which the (no longer) ``anomalous" features can be analytically derived.\\

\textbf{\textit{Quasicrystals}} -- Quasicrystals are crystalline structures possessing long range orientational order but lacking full translational periodicity \cite{steinhardt2019second,janssen1988aperiodic,divincenzo1999quasicrystals,janot1997quasicrystals}. Differently from glasses or amorphous systems, they display sharp Bragg peaks easily observable with X-ray diffraction. We refer to \cite{janssen2007aperiodic,stadnik2012physical,fan2016mathematical,scott2012incommensurate,jaric1988introduction} for a more extensive literature. The simplest examples are Penrose tilings and incommensurate structures. From a theoretical point of view, the best way to describe them is by using an extra-dimensional superspace formalism based on the mathematical fact that any aperiodic structure in $d$ dimensions is equivalent to a periodic one in a $D>d$ \textcolor{red}{manifold} cut at an irrational angle $\alpha$ \cite{janssen2014aperiodic}. In other words, to describe the crystalline structure of aperiodic crystals one needs a number of independent wave-vectors $D$ larger than the real spacetime dimensions $d$. This extra dynamics can be attributed to the so-called \textit{phason mode} \cite{C2CS35212E,Janssen2002,PhysRevB.34.3345,DEBOISSIEU2008107,PhysRevLett.54.1517,widom2008discussion,PhysRevLett.49.1833}. In the superspace formalism the phason simply corresponds to the internal shifts of the cut within the extra-dimensional \textcolor{red}{manifold}.\\

\begin{figure}
    \centering
    \includegraphics[width=0.6\linewidth]{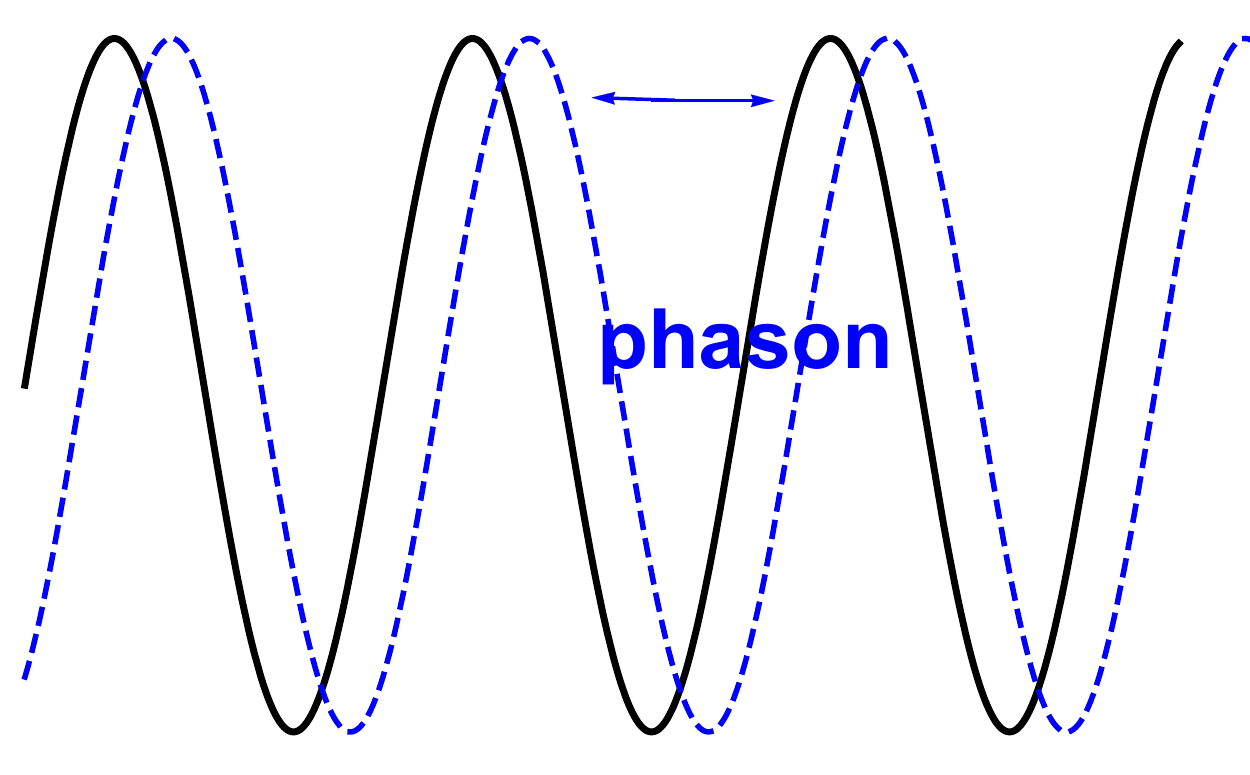}\\
     \includegraphics[width=0.6\linewidth]{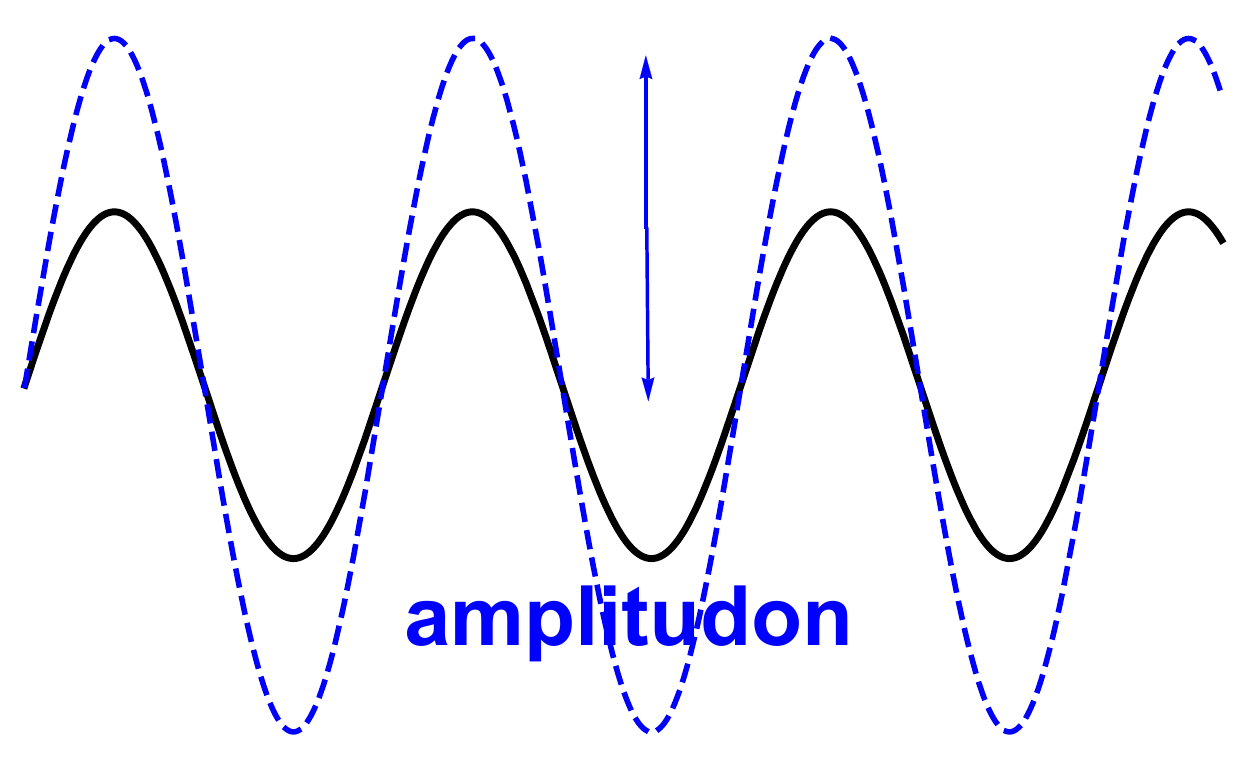}
    \caption{The two types of excitations in incommensurate structures. The phason, a shift in the phase of the modulated structure and the amplitudon, a variation in its amplitude.}
    \label{fig1}
\end{figure}

As a concrete case, let us focus on the dynamics of a modulated structure whose density is described as:
\begin{equation}
    \rho(r)\,=\,\rho_0\,+\,R\,\cos\left(k \,r+\phi(r)\right)
\end{equation}
where $k$ is the wave-vector of the substrate periodic lattice (e.g. the ionic lattice) and $\phi$ the phase of the modulation. Two possible excitations are possible:
\begin{equation}
    \rho(r)\,=\,\rho_0\,+\,\left(R+\delta R(r)\right)\,\cos\left(k \,r+\phi(r)+\delta \phi(r)\right)
\end{equation}
where $\delta R$ corresponds to the fluctuation of the amplitude -- \textit{amplitudon} -- and $\delta \phi$ to that of the phase -- \textit{phason}, (see Fig.\ref{fig1} for a visual representation). We refer to \cite{doi:10.1146/annurev-conmatphys-031214-014350} for a review of the role of these two types of excitation in condensed matter.\\

The crucial point is that, because of the lack of periodicity, there is an extra hydrodynamic ($\equiv$ massless) mode, the phason. Even more importantly, in the limit of small wave-vectors, its dispersion relation is \textit{diffusive}, as confirmed by several experimental checks \cite{PhysRevLett.91.225501,durand1991investigation}. Using a standard hydrodynamic formalism \cite{PhysRevB.32.7444}, the low energy dynamics of the phason is given by the dynamical equation:
\begin{equation}
    \partial_t\,\mathcal{Q}_{\phi}\,+\,\gamma\,\mathcal{Q}_{\phi}\,=\,-\frac{\partial \mathcal{F}}{\partial \phi}\,, \label{din}
\end{equation}
where $\phi$ is the phason shift, $\mathcal{Q}_{\phi}$ the conjugate momentum and $\mathcal{F}$ the free energy that now depends both on the phonon and phason shifts (see \cite{PhysRevB.32.7444} for details). The parameter $\gamma$ is the phason damping and the r.h.s. of the equation above is responsible for the propagative contribution in terms of the new phason \textcolor{red}{(generalized)} elastic moduli. Eq.\eqref{din} has exactly the same dynamics that one would find in a simple linear Maxwell model for viscoelasticity. Not surprisingly, the phason shows a k-gap crossover, which is postulated also for transverse phonons in liquids \cite{Baggioli:2019jcm}\footnote{Notice the difference with an attenuated sound mode following $\omega^2\,+\,i\,\Gamma\,\omega\,k^2=v^2k^2$.}. The phason damping is due to the fact that phason shifts are symmetries of the free energy but they do not commute with the Hamiltonian \cite{PhysRevLett.49.1833}. Technically, they are symmetries with no associated Noether current \cite{toappear}; physically, phason shifts cost energy because they correspond to atomic jumps and flips of the aperiodic structure (in analogy to atomic re-arrangements in liquids).\\ All in all, the solution of \eqref{din} can be written as
\begin{equation}
    \omega^2\,+\,i\,\omega\,\gamma\,=\,v^2\,k^2\,+\,\dots\,,
\end{equation}
which, for small momenta, predicts a diffusive behaviour
\begin{equation}
    \omega\,=\,-\,i\,D\,k^2\,+\,\dots;\quad D\,=\,v^2/\gamma\,. \label{usa}
\end{equation} 
The same result can be obtained from Frenkel-Kontorova models upon introducing a dissipative coefficient for relative motion between the two incommensurate superimposed lattices~\cite{chaikin2000principles}. 
At larger momenta, the phason recovers the standard propagative behaviour \cite{PhysRevLett.49.1833}, and can be thought of simply as an additional sound mode.
\begin{figure}
    \centering
    \includegraphics[width=0.9\linewidth]{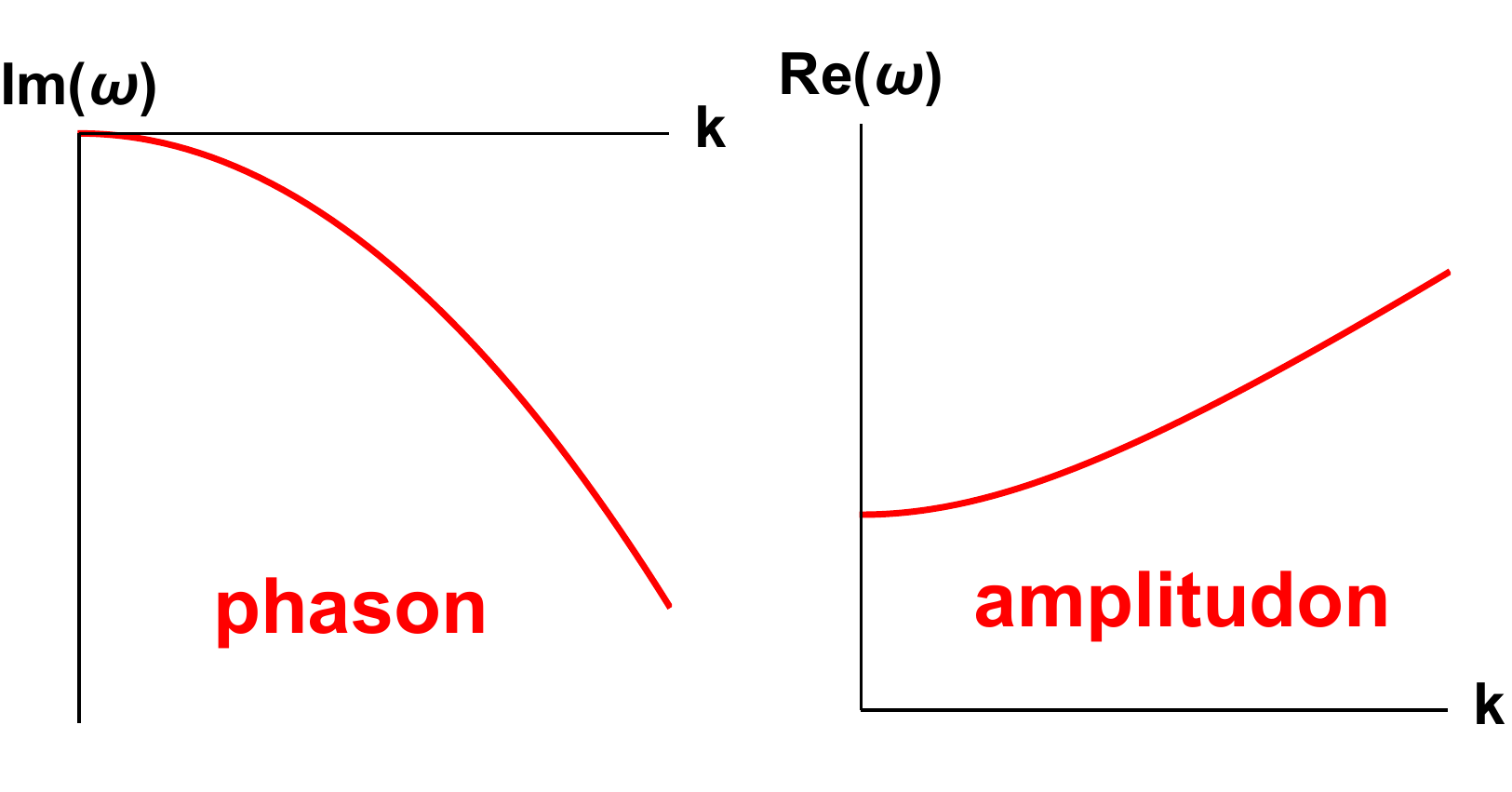}
    \caption{The low-energy dispersion relation of the phason and amplitudon excitations. The phason is a diffusive mode. The amplitudon is a gapped mode.}
    \label{fig2}
\end{figure}
\vspace{0.2cm}\\
On the contrary, the amplitudon is not a hydrodynamic gapless mode but rather a massive gapped excitation with dispersion relation:
\begin{equation}
    \omega^2\,=\,\omega_0^2\,+\,v^2\,k^2\,+\dots
\end{equation}
In a sense,  amplitudons can be thought of as additional optical-like phonon modes with the subtle difference that optical phonons do not have a clear interpretation in terms of spontaneous symmetry breaking (SSB)\footnote{Unless the optical-like phonons come from the interaction of layered structures \cite{Nika_2017} (in this context they are usually called pseudo-acoustic modes \cite{Esposito:2020hwq}).}. The amplitude mode is usually labelled as a ``Higgs mode'', in analogy to the Higgs particle, and its massive nature comes entirely from the SSB mechanism (e.g. amplitudons in superconducting transitions \cite{doi:10.1146/annurev-conmatphys-031119-050813}). \\

\textbf{\textit{Specific Heat}} --Interestingly, a robust linear in $T$ contribution to the low-temperature specific heat has been observed in incommensurate structures \cite{PhysRevLett.114.195502}. Clearly, such feature cannot be attributed to the presence of two-level states, but it was rather connected to the dynamics of the phason mode.  More specifically, a phenomenological computation \`{a} la Landau \cite{PhysRevLett.93.245902} found that:
\begin{equation}
    C\,\sim\,\frac{\gamma\,k_D}{v^2}\,T\,, \label{eq1}
\end{equation}
where $\gamma$ is the phason relaxation rate, $k_D$ the Debye momentum and $v$ the asymptotic phason speed.\\

Here, we show that such contribution is the typical and universal term appearing because of diffusive low energy modes due to the underlying fundamental symmetries. Using Eq.\eqref{usa}, the result \eqref{eq1} from \cite{PhysRevLett.93.245902} can be re-written as:
\begin{equation}
    C\,\sim\,\frac{k_D}{D}\,T\,,
\end{equation}
which is exactly the leading order term in \cite{PhysRevResearch.1.012010} (see Eqs. (10) and (13) therein). In Ref.\cite{PhysRevResearch.1.012010}, this result was obtained from consideration of ``diffusons'' ~\cite{allen1999diffusons} as the relevant excitations which manifest as a result of structural disorder in glasses, but share exactly the same mathematical form with \eqref{usa} for phasons above.

The result of \cite{PhysRevResearch.1.012010} is a first-principles analytic, and much simpler, computation which goes beyond the results of \cite{PhysRevLett.93.245902} and it assumes solely the existence of a diffusive low energy mode (in this case the phason).\\

This suggests that the ''anomalous'' linear in $T$ scaling of the specific heat in quasicrystals is due to the diffusive nature of the phason and can be understood in the general picture of \cite{PhysRevResearch.1.012010}. In this sense, phasons in quasycristal/incommensurate structures play exactly the same role as the diffusons in glasses and disordered materials \cite{allen1999diffusons}. Given the modest experimental evidence for TLS in glasses, and their non-applicability to quasicrystals, we propose that the presence of low-energy diffusive mode could be the real universal mechanism behind the (not so) anomalous linear in $T$ specific heat.\\

\begin{figure}[t]
    \centering
    \includegraphics[width=0.8 \linewidth]{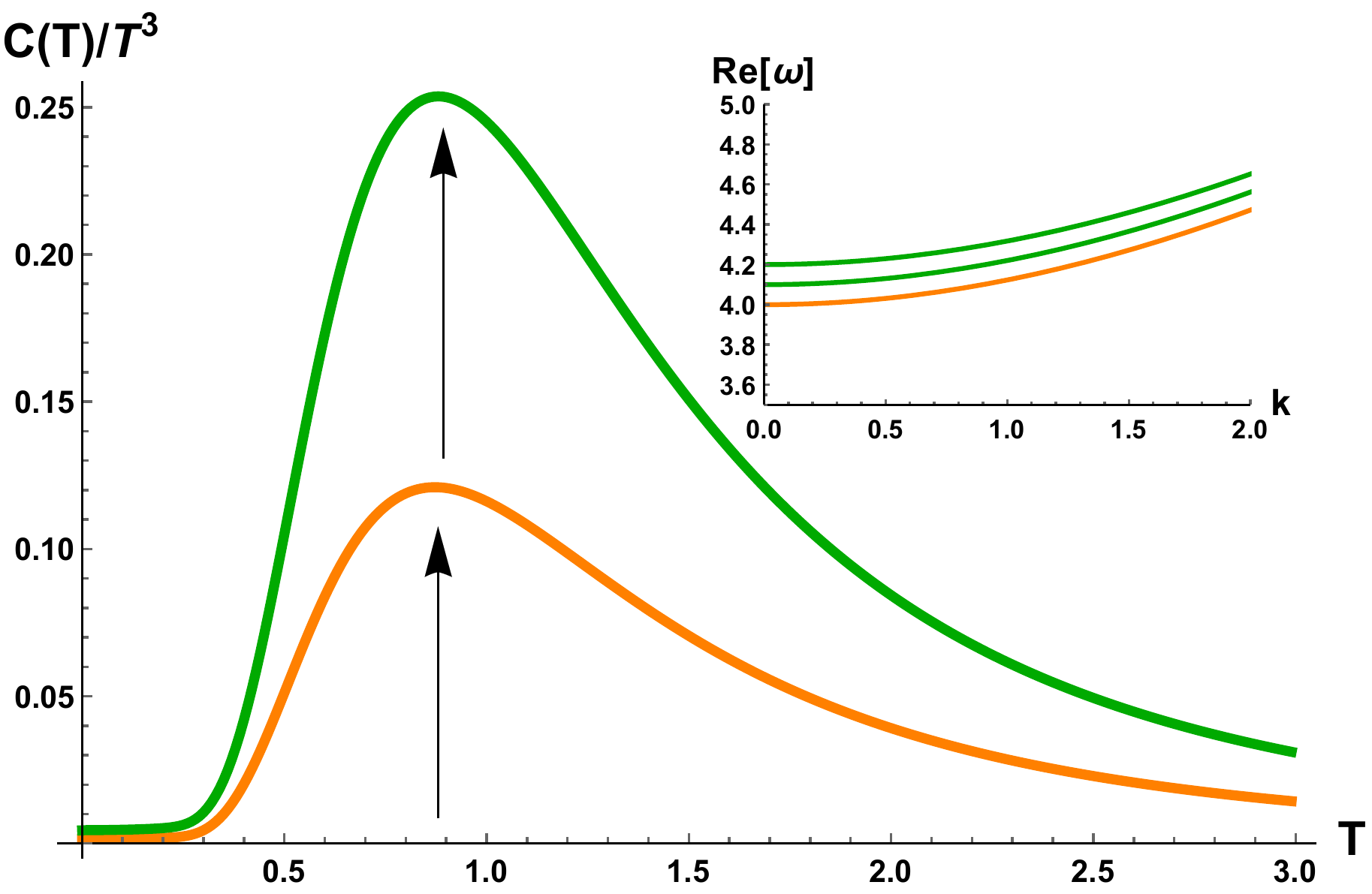}
    \caption{The contribution of gapped optical-like modes to the Debye-normalized specific heat. The orange curve is obtained using a single gapped mode; the green one by adding up three gapped modes. The inset shows the dispersion relation of the modes.}
    \label{fig:peak}
\end{figure}
In addition to the linear in $T$ contribution to the specific heat, a BP excess is experimentally observed in incommensurate structures as well \cite{PhysRevLett.114.195502}. The origin of this anomalous bump has been already attributed to the presence of gapped amplitudon modes \cite{PhysRevB.70.212301}. The similarities between this mechanism and those at work in ordered crystals \cite{Moratalla} went unnoticed until now. The resemblances are striking. It has been shown from direct measurements of the density of states and the low-T specific heat that the BP anomaly in ordered crystals \cite{Moratalla} is caused by the piling up of optical modes with a soft energy gap. \\

Both the observations, in quasicrystals and in ordered crystals, point towards the confirmation that gapped optical-like modes are able to produce a well-defined BP excess without the need of any structural disorder. These experimental facts are backed up by a first principle theoretical computation \cite{Baggioli_2019} which is in perfect agreement with the experimental results presented in both cases.\\

The idea is very simple and again based on symmetries and the low-energy dispersion relation of the modes. Take an underdamped gapped mode whose dispersion relation is well approximated by the solution of:
\begin{equation}
    \omega^2\,-\,\omega_0^2\,-\,v^2\,k^2\,+\,i\,\Gamma\,\omega\,=\,0\,, \label{gg}
\end{equation}
where $\omega_0$ is the energy gap of the mode. The parameter $\Gamma$ is the (Klemens) damping of the optical mode and it is taken to be $\Gamma \ll \omega_0$ such that the gapped excitation is a well-defined quasiparticle. Eq.\eqref{gg} corresponds to a quasiparticle Green function of the form:
\begin{equation}
    \mathcal{G}(\omega,k)\,=\,\frac{1}{\omega^2\,-\,\omega_0^2\,-\,v^2\,k^2\,+\,i\,\Gamma\,\omega}\,.
\end{equation}
Using the spectral representation of the density of states:
\begin{equation}
    g(\omega)\,=\,\frac{2\,\omega}{\pi\,k_D^3}\,\mathrm{Im}\,\int_0^{k_D}\,\mathcal{G}(\omega,k)\,d^3k
\end{equation}
together with the standard formula for the specific heat:
\begin{equation}
C(T)\,=\,k_B\,\int_0^\infty\, \left(\frac{\hbar \omega}{2\,k_B\,T}\right)^2\,\sinh \left(\frac{\hbar \omega}{2\,k_B\,T}\right)^{-2}\,g(\omega)\,d\omega\,,
\end{equation}
it is straightforward to compute the contribution of the gapped optical-like modes in Eq.\eqref{gg}.\\

The result is shown in Fig.\eqref{fig:peak} for an arbitrary choice of parameters. Importantly, the contribution of the gapped modes gives a BP excess whose location is mostly governed by the energy gap of the mode. Moreover, it is clear that piling up a large number of close-by gapped modes increases the amplitude of the BP and amplifies the effect. This simple computation provides a clear theoretical confirmation for the experimental results found in ordered crystals \cite{Moratalla} and quasicrystals \cite{PhysRevLett.114.195502}. Once again, the similar physical behaviour obtained in two completely different systems calls for a universal mechanism, which can be found in the gapped nature of the low-energy optical-like modes.\\

\textbf{\textit{Discussion}} --In this Letter, we demystify the nature of anomalous glassy features in the low-$T$ properties of solids. We collected experimental and theoretical evidence which supports the following emerging paradigm: (I) the anomalous thermal properties of glasses are not peculiar but can be found in different structures such as ordered crystals and incommensurate systems. In this sense, the label ``anomalous'' becomes meaningless. (II) Disorder (and all the theoretical models which incarnates it) and the presence of ``randomly distributed two-level systems'' cannot be the universal causes behind this much more general physics. They certainly cannot account for these features in ordered systems and in incommensurate systems.\\

We propose that a universal mechanism behind the linear in $T$ specific heat can be identified with the existence of low-energy quasi-localized diffusive modes. The existence of these modes can be caused by diverse physical features such as disorder-induced scattering (e.g. diffusons in glasses) or new low-energy dynamics due to incommensuration and aperiodicity (e.g. phasons in quasicrystals). We are able to prove analytically that low-energy diffusive modes give a linear in $T$ contribution to the specific heat, whose leading term is in perfect agreement with the phenomenological analysis of \cite{PhysRevLett.93.245902}.\\

In the same spirit, we argue that the presence of a boson peak anomaly can be generically induced by strong Akhiezer damping effects (due to high anharmonicities) and/or by the piling up of soft optical-like modes. 
The first mechanism could be responsible for the BP observed in the low-T specific heat of molecular crystals and cryocrystals~\cite{Krivchikov}, where anharmonicity is strong due to the shallow attractive part of the (van der Waals-type) interaction potential, thus leading to strong Akhiezer damping, which implies a diffusive mode, and hence to a BP~\cite{PhysRevLett.122.145501}.
The second mechanism is the one at work in certain ordered crystals, where the BP has been successfully linked to low-lying optical phonons \cite{PhysRevB.99.024301} and quasicrystals, where the gapped modes are the amplitude fluctuations of the incommensurate structure \cite{PhysRevB.70.212301}.\\

\begin{figure}[t!]
    \centering
    \includegraphics[width=0.7\linewidth]{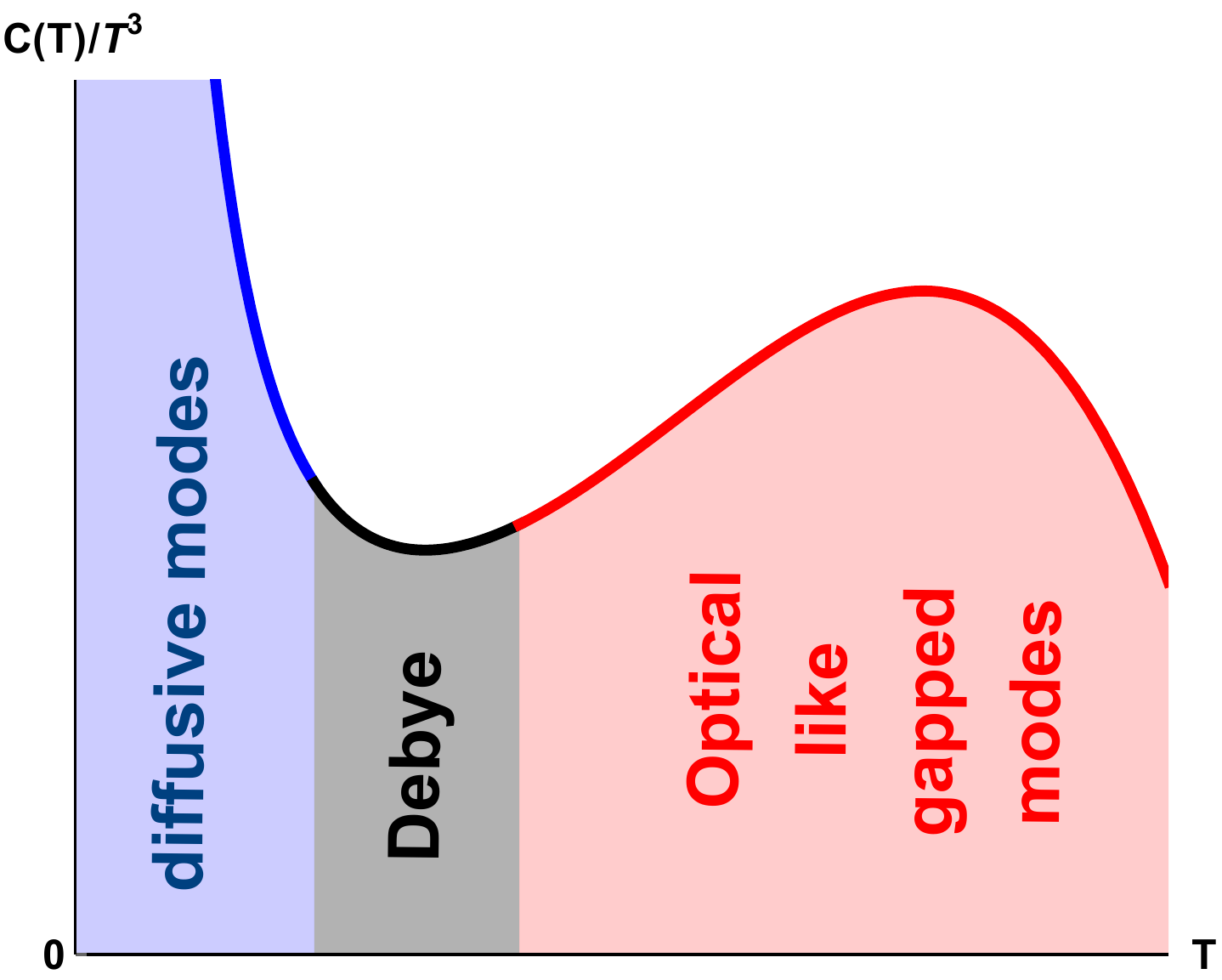}
    \caption{The typical structure of the Debye-normalized specific heat in glasses, ordered crystals and incommensurate structures. At low temperature, a $\sim 1/T^2$ fall-off is visible before the Debye regime $\sim cost$. Finally, a BP bump is observed at higher temperatures. In our interpretation, all these features can be universally explained by the presence of quasi-localized diffusive modes and soft optical-like gapped modes.}
    \label{fig3}
\end{figure}

Given these interesting formal analogies between diffusive excitations in glasses and aperiodic crystals, it is likely that the same low-temperature anomalies could be found in the thermal conductivity $\kappa$. Not surprisingly, a $\sim T^2$ scaling and a glass-like plateau have already been observed in the thermal conductivity of aperiodic crystals \cite{PhysRevB.51.153,doi:10.1002/ijch.201100150}. It is tempting to attribute them to the existence of diffusive phason modes playing the same role of diffusons~\cite{allen1999diffusons} in glasses. This would provide a final confirmation of the universal picture presented in this work.\\

Moreover, phasons appear also in twisted bilayer graphene (TBG), where they might play a key role for transport and thermodynamics \cite{PhysRevB.100.155426}. Following the same logic, it is tempting to predict the presence of a linear in $T$ term in the vibrational specific heat of TBG. To the best of our knowledge, measurements of the specific heat in TBG have not appeared yet.\\

In summary, we suggest that the universal (and not anymore anomalous) glassy-like features have to be generically understood from the nature of the low-energy excitations (see Fig.\ref{fig3} for a visual summary). The latter is the key behind these, more ubiquitous than originally thought, features, which importantly do not depend exclusively on the presence of disorder or any amorphous structure. Supported by experimental and theoretical evidence, this view sheds new light onto an old problem stuck into old ideas.\\

\textbf{\textit{Acknowledgments}} -- We thank A.Cano, F.Guinea, A.I.Krivchikov, M.Landry, H.Ochoa, M.A. Ramos and J.L. Tamarit for continuous and enlighting discussions about these topics. M.B. acknowledges the support of the Spanish MINECO’s ``Centro de Excelencia Severo Ochoa'' Programme under grant SEV-2012-0249.
\bibliographystyle{apsrev4-1}

\bibliography{phason}

%merlin.mbs apsrev4-1.bst 2010-07-25 4.21a (PWD, AO, DPC) hacked
%Control: key (0)
%Control: author (72) initials jnrlst
%Control: editor formatted (1) identically to author
%Control: production of article title (-1) disabled
%Control: page (0) single
%Control: year (1) truncated
%Control: production of eprint (0) enabled
\begin{thebibliography}{57}%
\makeatletter
\providecommand \@ifxundefined [1]{%
 \@ifx{#1\undefined}
}%
\providecommand \@ifnum [1]{%
 \ifnum #1\expandafter \@firstoftwo
 \else \expandafter \@secondoftwo
 \fi
}%
\providecommand \@ifx [1]{%
 \ifx #1\expandafter \@firstoftwo
 \else \expandafter \@secondoftwo
 \fi
}%
\providecommand \natexlab [1]{#1}%
\providecommand \enquote  [1]{``#1''}%
\providecommand \bibnamefont  [1]{#1}%
\providecommand \bibfnamefont [1]{#1}%
\providecommand \citenamefont [1]{#1}%
\providecommand \href@noop [0]{\@secondoftwo}%
\providecommand \href [0]{\begingroup \@sanitize@url \@href}%
\providecommand \@href[1]{\@@startlink{#1}\@@href}%
\providecommand \@@href[1]{\endgroup#1\@@endlink}%
\providecommand \@sanitize@url [0]{\catcode `\\12\catcode `\$12\catcode
  `\&12\catcode `\#12\catcode `\^12\catcode `\_12\catcode `\%12\relax}%
\providecommand \@@startlink[1]{}%
\providecommand \@@endlink[0]{}%
\providecommand \url  [0]{\begingroup\@sanitize@url \@url }%
\providecommand \@url [1]{\endgroup\@href {#1}{\urlprefix }}%
\providecommand \urlprefix  [0]{URL }%
\providecommand \Eprint [0]{\href }%
\providecommand \doibase [0]{http://dx.doi.org/}%
\providecommand \selectlanguage [0]{\@gobble}%
\providecommand \bibinfo  [0]{\@secondoftwo}%
\providecommand \bibfield  [0]{\@secondoftwo}%
\providecommand \translation [1]{[#1]}%
\providecommand \BibitemOpen [0]{}%
\providecommand \bibitemStop [0]{}%
\providecommand \bibitemNoStop [0]{.\EOS\space}%
\providecommand \EOS [0]{\spacefactor3000\relax}%
\providecommand \BibitemShut  [1]{\csname bibitem#1\endcsname}%
\let\auto@bib@innerbib\@empty
%</preamble>
\bibitem [{\citenamefont {Zeller}\ and\ \citenamefont
  {Pohl}(1971)}]{PhysRevB.4.2029}%
  \BibitemOpen
  \bibfield  {author} {\bibinfo {author} {\bibfnamefont {R.~C.}\ \bibnamefont
  {Zeller}}\ and\ \bibinfo {author} {\bibfnamefont {R.~O.}\ \bibnamefont
  {Pohl}},\ }\href {\doibase 10.1103/PhysRevB.4.2029} {\bibfield  {journal}
  {\bibinfo  {journal} {Phys. Rev. B}\ }\textbf {\bibinfo {volume} {4}},\
  \bibinfo {pages} {2029} (\bibinfo {year} {1971})}\BibitemShut {NoStop}%
\bibitem [{\citenamefont {Shintani}\ and\ \citenamefont
  {Tanaka}(2008)}]{Shintani2008}%
  \BibitemOpen
  \bibfield  {author} {\bibinfo {author} {\bibfnamefont {H.}~\bibnamefont
  {Shintani}}\ and\ \bibinfo {author} {\bibfnamefont {H.}~\bibnamefont
  {Tanaka}},\ }\href {\doibase 10.1038/nmat2293} {\bibfield  {journal}
  {\bibinfo  {journal} {Nature Materials}\ }\textbf {\bibinfo {volume} {7}},\
  \bibinfo {pages} {870} (\bibinfo {year} {2008})}\BibitemShut {NoStop}%
\bibitem [{\citenamefont {Beltukov}\ \emph {et~al.}(2013)\citenamefont
  {Beltukov}, \citenamefont {Kozub},\ and\ \citenamefont
  {Parshin}}]{PhysRevB.87.134203}%
  \BibitemOpen
  \bibfield  {author} {\bibinfo {author} {\bibfnamefont {Y.~M.}\ \bibnamefont
  {Beltukov}}, \bibinfo {author} {\bibfnamefont {V.~I.}\ \bibnamefont {Kozub}},
  \ and\ \bibinfo {author} {\bibfnamefont {D.~A.}\ \bibnamefont {Parshin}},\
  }\href {\doibase 10.1103/PhysRevB.87.134203} {\bibfield  {journal} {\bibinfo
  {journal} {Phys. Rev. B}\ }\textbf {\bibinfo {volume} {87}},\ \bibinfo
  {pages} {134203} (\bibinfo {year} {2013})}\BibitemShut {NoStop}%
\bibitem [{\citenamefont {Ruffl\'e}\ \emph {et~al.}(2006)\citenamefont
  {Ruffl\'e}, \citenamefont {Guimbreti\`ere}, \citenamefont {Courtens},
  \citenamefont {Vacher},\ and\ \citenamefont
  {Monaco}}]{PhysRevLett.96.045502}%
  \BibitemOpen
  \bibfield  {author} {\bibinfo {author} {\bibfnamefont {B.}~\bibnamefont
  {Ruffl\'e}}, \bibinfo {author} {\bibfnamefont {G.}~\bibnamefont
  {Guimbreti\`ere}}, \bibinfo {author} {\bibfnamefont {E.}~\bibnamefont
  {Courtens}}, \bibinfo {author} {\bibfnamefont {R.}~\bibnamefont {Vacher}}, \
  and\ \bibinfo {author} {\bibfnamefont {G.}~\bibnamefont {Monaco}},\ }\href
  {\doibase 10.1103/PhysRevLett.96.045502} {\bibfield  {journal} {\bibinfo
  {journal} {Phys. Rev. Lett.}\ }\textbf {\bibinfo {volume} {96}},\ \bibinfo
  {pages} {045502} (\bibinfo {year} {2006})}\BibitemShut {NoStop}%
\bibitem [{\citenamefont {Phillips}(1972)}]{Phillips1972}%
  \BibitemOpen
  \bibfield  {author} {\bibinfo {author} {\bibfnamefont {W.~A.}\ \bibnamefont
  {Phillips}},\ }\href {\doibase 10.1007/BF00660072} {\bibfield  {journal}
  {\bibinfo  {journal} {Journal of Low Temperature Physics}\ }\textbf {\bibinfo
  {volume} {7}},\ \bibinfo {pages} {351} (\bibinfo {year} {1972})}\BibitemShut
  {NoStop}%
\bibitem [{\citenamefont {Anderson}\ \emph {et~al.}(1972)\citenamefont
  {Anderson}, \citenamefont {Halperin},\ and\ \citenamefont
  {Varma}}]{Anderson1972-ANDALT-2}%
  \BibitemOpen
  \bibfield  {author} {\bibinfo {author} {\bibfnamefont {P.~W.}\ \bibnamefont
  {Anderson}}, \bibinfo {author} {\bibfnamefont {B.~I.}\ \bibnamefont
  {Halperin}}, \ and\ \bibinfo {author} {\bibfnamefont {C.~M.}\ \bibnamefont
  {Varma}},\ }\href {\doibase 10.1080/14786437208229210} {\bibfield  {journal}
  {\bibinfo  {journal} {Philosophical Magazine}\ }\textbf {\bibinfo {volume}
  {25}},\ \bibinfo {pages} {1} (\bibinfo {year} {1972})}\BibitemShut {NoStop}%
\bibitem [{\citenamefont {{Phillips}}(1987)}]{1987RPPh...50.1657P}%
  \BibitemOpen
  \bibfield  {author} {\bibinfo {author} {\bibfnamefont {W.~A.}\ \bibnamefont
  {{Phillips}}},\ }\href {\doibase 10.1088/0034-4885/50/12/003} {\bibfield
  {journal} {\bibinfo  {journal} {Reports on Progress in Physics}\ }\textbf
  {\bibinfo {volume} {50}},\ \bibinfo {pages} {1657} (\bibinfo {year}
  {1987})}\BibitemShut {NoStop}%
\bibitem [{\citenamefont {Schirmacher}\ \emph {et~al.}(2007)\citenamefont
  {Schirmacher}, \citenamefont {Ruocco},\ and\ \citenamefont
  {Scopigno}}]{Ruocco2007}%
  \BibitemOpen
  \bibfield  {author} {\bibinfo {author} {\bibfnamefont {W.}~\bibnamefont
  {Schirmacher}}, \bibinfo {author} {\bibfnamefont {G.}~\bibnamefont {Ruocco}},
  \ and\ \bibinfo {author} {\bibfnamefont {T.}~\bibnamefont {Scopigno}},\
  }\href {\doibase 10.1103/PhysRevLett.98.025501} {\bibfield  {journal}
  {\bibinfo  {journal} {Phys. Rev. Lett.}\ }\textbf {\bibinfo {volume} {98}},\
  \bibinfo {pages} {025501} (\bibinfo {year} {2007})}\BibitemShut {NoStop}%
\bibitem [{\citenamefont {Taraskin}\ \emph {et~al.}(2001)\citenamefont
  {Taraskin}, \citenamefont {Loh}, \citenamefont {Natarajan},\ and\
  \citenamefont {Elliott}}]{Taraskin}%
  \BibitemOpen
  \bibfield  {author} {\bibinfo {author} {\bibfnamefont {S.~N.}\ \bibnamefont
  {Taraskin}}, \bibinfo {author} {\bibfnamefont {Y.~L.}\ \bibnamefont {Loh}},
  \bibinfo {author} {\bibfnamefont {G.}~\bibnamefont {Natarajan}}, \ and\
  \bibinfo {author} {\bibfnamefont {S.~R.}\ \bibnamefont {Elliott}},\ }\href
  {\doibase 10.1103/PhysRevLett.86.1255} {\bibfield  {journal} {\bibinfo
  {journal} {Phys. Rev. Lett.}\ }\textbf {\bibinfo {volume} {86}},\ \bibinfo
  {pages} {1255} (\bibinfo {year} {2001})}\BibitemShut {NoStop}%
\bibitem [{\citenamefont {Buchenau}\ \emph {et~al.}(1991)\citenamefont
  {Buchenau}, \citenamefont {Galperin}, \citenamefont {Gurevich},\ and\
  \citenamefont {Schober}}]{PhysRevB.43.5039}%
  \BibitemOpen
  \bibfield  {author} {\bibinfo {author} {\bibfnamefont {U.}~\bibnamefont
  {Buchenau}}, \bibinfo {author} {\bibfnamefont {Y.~M.}\ \bibnamefont
  {Galperin}}, \bibinfo {author} {\bibfnamefont {V.~L.}\ \bibnamefont
  {Gurevich}}, \ and\ \bibinfo {author} {\bibfnamefont {H.~R.}\ \bibnamefont
  {Schober}},\ }\href {\doibase 10.1103/PhysRevB.43.5039} {\bibfield  {journal}
  {\bibinfo  {journal} {Phys. Rev. B}\ }\textbf {\bibinfo {volume} {43}},\
  \bibinfo {pages} {5039} (\bibinfo {year} {1991})}\BibitemShut {NoStop}%
\bibitem [{\citenamefont {Leggett}\ and\ \citenamefont
  {Vural}(2013)}]{doi:10.1021/jp402222g}%
  \BibitemOpen
  \bibfield  {author} {\bibinfo {author} {\bibfnamefont {A.~J.}\ \bibnamefont
  {Leggett}}\ and\ \bibinfo {author} {\bibfnamefont {D.~C.}\ \bibnamefont
  {Vural}},\ }\href {\doibase 10.1021/jp402222g} {\bibfield  {journal}
  {\bibinfo  {journal} {The Journal of Physical Chemistry B}\ }\textbf
  {\bibinfo {volume} {117}},\ \bibinfo {pages} {12966} (\bibinfo {year}
  {2013})},\ \bibinfo {note} {pMID: 23924397},\ \Eprint
  {http://arxiv.org/abs/https://doi.org/10.1021/jp402222g}
  {https://doi.org/10.1021/jp402222g} \BibitemShut {NoStop}%
\bibitem [{\citenamefont {Ramos}(2020)}]{ramos2020universal}%
  \BibitemOpen
  \bibfield  {author} {\bibinfo {author} {\bibfnamefont {M.~A.}\ \bibnamefont
  {Ramos}},\ }\href@noop {} {\bibfield  {journal} {\bibinfo  {journal} {Low
  Temperature Physics}\ }\textbf {\bibinfo {volume} {46}},\ \bibinfo {pages}
  {104} (\bibinfo {year} {2020})}\BibitemShut {NoStop}%
\bibitem [{\citenamefont {Strzhemechny}\ \emph {et~al.}(2019)\citenamefont
  {Strzhemechny}, \citenamefont {Krivchikov},\ and\ \citenamefont
  {Jeżowski}}]{Krivchikov}%
  \BibitemOpen
  \bibfield  {author} {\bibinfo {author} {\bibfnamefont {M.~A.}\ \bibnamefont
  {Strzhemechny}}, \bibinfo {author} {\bibfnamefont {A.~I.}\ \bibnamefont
  {Krivchikov}}, \ and\ \bibinfo {author} {\bibfnamefont {A.}~\bibnamefont
  {Jeżowski}},\ }\href {\doibase 10.1063/10.0000211} {\bibfield  {journal}
  {\bibinfo  {journal} {Low Temperature Physics}\ }\textbf {\bibinfo {volume}
  {45}},\ \bibinfo {pages} {1290} (\bibinfo {year} {2019})},\ \Eprint
  {http://arxiv.org/abs/https://doi.org/10.1063/10.0000211}
  {https://doi.org/10.1063/10.0000211} \BibitemShut {NoStop}%
\bibitem [{\citenamefont {Gebbia}\ \emph {et~al.}(2017)\citenamefont {Gebbia},
  \citenamefont {Ramos}, \citenamefont {Szewczyk}, \citenamefont {Jezowski},
  \citenamefont {Krivchikov}, \citenamefont {Horbatenko}, \citenamefont
  {Guidi}, \citenamefont {Bermejo},\ and\ \citenamefont
  {Tamarit}}]{PhysRevLett.119.215506}%
  \BibitemOpen
  \bibfield  {author} {\bibinfo {author} {\bibfnamefont {J.~F.}\ \bibnamefont
  {Gebbia}}, \bibinfo {author} {\bibfnamefont {M.~A.}\ \bibnamefont {Ramos}},
  \bibinfo {author} {\bibfnamefont {D.}~\bibnamefont {Szewczyk}}, \bibinfo
  {author} {\bibfnamefont {A.}~\bibnamefont {Jezowski}}, \bibinfo {author}
  {\bibfnamefont {A.~I.}\ \bibnamefont {Krivchikov}}, \bibinfo {author}
  {\bibfnamefont {Y.~V.}\ \bibnamefont {Horbatenko}}, \bibinfo {author}
  {\bibfnamefont {T.}~\bibnamefont {Guidi}}, \bibinfo {author} {\bibfnamefont
  {F.~J.}\ \bibnamefont {Bermejo}}, \ and\ \bibinfo {author} {\bibfnamefont
  {J.~L.}\ \bibnamefont {Tamarit}},\ }\href {\doibase
  10.1103/PhysRevLett.119.215506} {\bibfield  {journal} {\bibinfo  {journal}
  {Phys. Rev. Lett.}\ }\textbf {\bibinfo {volume} {119}},\ \bibinfo {pages}
  {215506} (\bibinfo {year} {2017})}\BibitemShut {NoStop}%
\bibitem [{\citenamefont {Moratalla}\ \emph
  {et~al.}(2019{\natexlab{a}})\citenamefont {Moratalla}, \citenamefont
  {Gebbia}, \citenamefont {Ramos}, \citenamefont {Pardo}, \citenamefont
  {Mukhopadhyay}, \citenamefont {Rudi\ifmmode~\acute{c}\else \'{c}\fi{}},
  \citenamefont {Fernandez-Alonso}, \citenamefont {Bermejo},\ and\
  \citenamefont {Tamarit}}]{PhysRevB.99.024301}%
  \BibitemOpen
  \bibfield  {author} {\bibinfo {author} {\bibfnamefont {M.}~\bibnamefont
  {Moratalla}}, \bibinfo {author} {\bibfnamefont {J.~F.}\ \bibnamefont
  {Gebbia}}, \bibinfo {author} {\bibfnamefont {M.~A.}\ \bibnamefont {Ramos}},
  \bibinfo {author} {\bibfnamefont {L.~C.}\ \bibnamefont {Pardo}}, \bibinfo
  {author} {\bibfnamefont {S.}~\bibnamefont {Mukhopadhyay}}, \bibinfo {author}
  {\bibfnamefont {S.}~\bibnamefont {Rudi\ifmmode~\acute{c}\else \'{c}\fi{}}},
  \bibinfo {author} {\bibfnamefont {F.}~\bibnamefont {Fernandez-Alonso}},
  \bibinfo {author} {\bibfnamefont {F.~J.}\ \bibnamefont {Bermejo}}, \ and\
  \bibinfo {author} {\bibfnamefont {J.~L.}\ \bibnamefont {Tamarit}},\ }\href
  {\doibase 10.1103/PhysRevB.99.024301} {\bibfield  {journal} {\bibinfo
  {journal} {Phys. Rev. B}\ }\textbf {\bibinfo {volume} {99}},\ \bibinfo
  {pages} {024301} (\bibinfo {year} {2019}{\natexlab{a}})}\BibitemShut
  {NoStop}%
\bibitem [{\citenamefont {Rem\'enyi}\ \emph {et~al.}(2015)\citenamefont
  {Rem\'enyi}, \citenamefont {Sahling}, \citenamefont
  {Biljakovi\ifmmode~\acute{c}\else \'{c}\fi{}}, \citenamefont {Stare\ifmmode
  \check{s}\else \v{s}\fi{}ini\ifmmode~\acute{c}\else \'{c}\fi{}},
  \citenamefont {Lasjaunias}, \citenamefont {Lorenzo}, \citenamefont
  {Monceau},\ and\ \citenamefont {Cano}}]{PhysRevLett.114.195502}%
  \BibitemOpen
  \bibfield  {author} {\bibinfo {author} {\bibfnamefont {G.}~\bibnamefont
  {Rem\'enyi}}, \bibinfo {author} {\bibfnamefont {S.}~\bibnamefont {Sahling}},
  \bibinfo {author} {\bibfnamefont {K.}~\bibnamefont
  {Biljakovi\ifmmode~\acute{c}\else \'{c}\fi{}}}, \bibinfo {author}
  {\bibfnamefont {D.}~\bibnamefont {Stare\ifmmode \check{s}\else
  \v{s}\fi{}ini\ifmmode~\acute{c}\else \'{c}\fi{}}}, \bibinfo {author}
  {\bibfnamefont {J.-C.}\ \bibnamefont {Lasjaunias}}, \bibinfo {author}
  {\bibfnamefont {J.~E.}\ \bibnamefont {Lorenzo}}, \bibinfo {author}
  {\bibfnamefont {P.}~\bibnamefont {Monceau}}, \ and\ \bibinfo {author}
  {\bibfnamefont {A.}~\bibnamefont {Cano}},\ }\href {\doibase
  10.1103/PhysRevLett.114.195502} {\bibfield  {journal} {\bibinfo  {journal}
  {Phys. Rev. Lett.}\ }\textbf {\bibinfo {volume} {114}},\ \bibinfo {pages}
  {195502} (\bibinfo {year} {2015})}\BibitemShut {NoStop}%
\bibitem [{\citenamefont {Baggioli}\ and\ \citenamefont
  {Zaccone}(2020)}]{PhysRevResearch.2.013267}%
  \BibitemOpen
  \bibfield  {author} {\bibinfo {author} {\bibfnamefont {M.}~\bibnamefont
  {Baggioli}}\ and\ \bibinfo {author} {\bibfnamefont {A.}~\bibnamefont
  {Zaccone}},\ }\href {\doibase 10.1103/PhysRevResearch.2.013267} {\bibfield
  {journal} {\bibinfo  {journal} {Phys. Rev. Research}\ }\textbf {\bibinfo
  {volume} {2}},\ \bibinfo {pages} {013267} (\bibinfo {year}
  {2020})}\BibitemShut {NoStop}%
\bibitem [{\citenamefont {Baggioli}\ and\ \citenamefont
  {Zaccone}(2019{\natexlab{a}})}]{PhysRevLett.122.145501}%
  \BibitemOpen
  \bibfield  {author} {\bibinfo {author} {\bibfnamefont {M.}~\bibnamefont
  {Baggioli}}\ and\ \bibinfo {author} {\bibfnamefont {A.}~\bibnamefont
  {Zaccone}},\ }\href {\doibase 10.1103/PhysRevLett.122.145501} {\bibfield
  {journal} {\bibinfo  {journal} {Phys. Rev. Lett.}\ }\textbf {\bibinfo
  {volume} {122}},\ \bibinfo {pages} {145501} (\bibinfo {year}
  {2019}{\natexlab{a}})}\BibitemShut {NoStop}%
\bibitem [{\citenamefont {Casella}\ \emph {et~al.}(2020)\citenamefont
  {Casella}, \citenamefont {Baggioli}, \citenamefont {Mori},\ and\
  \citenamefont {Zaccone}}]{casella2020physics}%
  \BibitemOpen
  \bibfield  {author} {\bibinfo {author} {\bibfnamefont {L.}~\bibnamefont
  {Casella}}, \bibinfo {author} {\bibfnamefont {M.}~\bibnamefont {Baggioli}},
  \bibinfo {author} {\bibfnamefont {T.}~\bibnamefont {Mori}}, \ and\ \bibinfo
  {author} {\bibfnamefont {A.}~\bibnamefont {Zaccone}},\ }\href@noop {}
  {\enquote {\bibinfo {title} {Physics of phonon-polaritons in amorphous
  materials},}\ } (\bibinfo {year} {2020}),\ \Eprint
  {http://arxiv.org/abs/2004.12358} {arXiv:2004.12358 [cond-mat.dis-nn]}
  \BibitemShut {NoStop}%
\bibitem [{\citenamefont {Baggioli}\ and\ \citenamefont
  {Zaccone}(2019{\natexlab{b}})}]{Baggioli_2019}%
  \BibitemOpen
  \bibfield  {author} {\bibinfo {author} {\bibfnamefont {M.}~\bibnamefont
  {Baggioli}}\ and\ \bibinfo {author} {\bibfnamefont {A.}~\bibnamefont
  {Zaccone}},\ }\href {\doibase 10.1088/2515-7639/ab4758} {\bibfield  {journal}
  {\bibinfo  {journal} {Journal of Physics: Materials}\ }\textbf {\bibinfo
  {volume} {3}},\ \bibinfo {pages} {015004} (\bibinfo {year}
  {2019}{\natexlab{b}})}\BibitemShut {NoStop}%
\bibitem [{\citenamefont {Baggioli}\ \emph
  {et~al.}(2019{\natexlab{a}})\citenamefont {Baggioli}, \citenamefont {Cui},\
  and\ \citenamefont {Zaccone}}]{PhysRevB.100.220201}%
  \BibitemOpen
  \bibfield  {author} {\bibinfo {author} {\bibfnamefont {M.}~\bibnamefont
  {Baggioli}}, \bibinfo {author} {\bibfnamefont {B.}~\bibnamefont {Cui}}, \
  and\ \bibinfo {author} {\bibfnamefont {A.}~\bibnamefont {Zaccone}},\ }\href
  {\doibase 10.1103/PhysRevB.100.220201} {\bibfield  {journal} {\bibinfo
  {journal} {Phys. Rev. B}\ }\textbf {\bibinfo {volume} {100}},\ \bibinfo
  {pages} {220201} (\bibinfo {year} {2019}{\natexlab{a}})}\BibitemShut
  {NoStop}%
\bibitem [{\citenamefont {Baggioli}\ and\ \citenamefont
  {Zaccone}(2019{\natexlab{c}})}]{PhysRevResearch.1.012010}%
  \BibitemOpen
  \bibfield  {author} {\bibinfo {author} {\bibfnamefont {M.}~\bibnamefont
  {Baggioli}}\ and\ \bibinfo {author} {\bibfnamefont {A.}~\bibnamefont
  {Zaccone}},\ }\href {\doibase 10.1103/PhysRevResearch.1.012010} {\bibfield
  {journal} {\bibinfo  {journal} {Phys. Rev. Research}\ }\textbf {\bibinfo
  {volume} {1}},\ \bibinfo {pages} {012010} (\bibinfo {year}
  {2019}{\natexlab{c}})}\BibitemShut {NoStop}%
\bibitem [{\citenamefont {Baggioli}\ \emph
  {et~al.}(2019{\natexlab{b}})\citenamefont {Baggioli}, \citenamefont
  {Milkus},\ and\ \citenamefont {Zaccone}}]{PhysRevE.100.062131}%
  \BibitemOpen
  \bibfield  {author} {\bibinfo {author} {\bibfnamefont {M.}~\bibnamefont
  {Baggioli}}, \bibinfo {author} {\bibfnamefont {R.}~\bibnamefont {Milkus}}, \
  and\ \bibinfo {author} {\bibfnamefont {A.}~\bibnamefont {Zaccone}},\ }\href
  {\doibase 10.1103/PhysRevE.100.062131} {\bibfield  {journal} {\bibinfo
  {journal} {Phys. Rev. E}\ }\textbf {\bibinfo {volume} {100}},\ \bibinfo
  {pages} {062131} (\bibinfo {year} {2019}{\natexlab{b}})}\BibitemShut
  {NoStop}%
\bibitem [{\citenamefont {Cano}\ and\ \citenamefont
  {Levanyuk}(2004{\natexlab{a}})}]{PhysRevLett.93.245902}%
  \BibitemOpen
  \bibfield  {author} {\bibinfo {author} {\bibfnamefont {A.}~\bibnamefont
  {Cano}}\ and\ \bibinfo {author} {\bibfnamefont {A.~P.}\ \bibnamefont
  {Levanyuk}},\ }\href {\doibase 10.1103/PhysRevLett.93.245902} {\bibfield
  {journal} {\bibinfo  {journal} {Phys. Rev. Lett.}\ }\textbf {\bibinfo
  {volume} {93}},\ \bibinfo {pages} {245902} (\bibinfo {year}
  {2004}{\natexlab{a}})}\BibitemShut {NoStop}%
\bibitem [{\citenamefont {Cano}\ and\ \citenamefont
  {Levanyuk}(2004{\natexlab{b}})}]{PhysRevB.70.212301}%
  \BibitemOpen
  \bibfield  {author} {\bibinfo {author} {\bibfnamefont {A.}~\bibnamefont
  {Cano}}\ and\ \bibinfo {author} {\bibfnamefont {A.~P.}\ \bibnamefont
  {Levanyuk}},\ }\href {\doibase 10.1103/PhysRevB.70.212301} {\bibfield
  {journal} {\bibinfo  {journal} {Phys. Rev. B}\ }\textbf {\bibinfo {volume}
  {70}},\ \bibinfo {pages} {212301} (\bibinfo {year}
  {2004}{\natexlab{b}})}\BibitemShut {NoStop}%
\bibitem [{\citenamefont {Steinhardt}(2019)}]{steinhardt2019second}%
  \BibitemOpen
  \bibfield  {author} {\bibinfo {author} {\bibfnamefont {P.}~\bibnamefont
  {Steinhardt}},\ }\href {https://books.google.es/books?id=GWBEDwAAQBAJ} {\emph
  {\bibinfo {title} {The Second Kind of Impossible: The Extraordinary Quest for
  a New Form of Matter}}}\ (\bibinfo  {publisher} {Simon \& Schuster},\
  \bibinfo {year} {2019})\BibitemShut {NoStop}%
\bibitem [{\citenamefont {Janssen}(1988)}]{janssen1988aperiodic}%
  \BibitemOpen
  \bibfield  {author} {\bibinfo {author} {\bibfnamefont {T.}~\bibnamefont
  {Janssen}},\ }\href@noop {} {\bibfield  {journal} {\bibinfo  {journal}
  {Physics Reports}\ }\textbf {\bibinfo {volume} {168}},\ \bibinfo {pages} {55}
  (\bibinfo {year} {1988})}\BibitemShut {NoStop}%
\bibitem [{\citenamefont {DiVincenzo}\ and\ \citenamefont
  {Steinhardt}(1999)}]{divincenzo1999quasicrystals}%
  \BibitemOpen
  \bibfield  {author} {\bibinfo {author} {\bibfnamefont {D.}~\bibnamefont
  {DiVincenzo}}\ and\ \bibinfo {author} {\bibfnamefont {P.}~\bibnamefont
  {Steinhardt}},\ }\href {https://books.google.es/books?id=Jk3WFRUslRYC} {\emph
  {\bibinfo {title} {Quasicrystals: The State of the Art}}},\ Series on
  directions in condensed matter physics\ (\bibinfo  {publisher} {World
  Scientific},\ \bibinfo {year} {1999})\BibitemShut {NoStop}%
\bibitem [{\citenamefont {Janot}(1997)}]{janot1997quasicrystals}%
  \BibitemOpen
  \bibfield  {author} {\bibinfo {author} {\bibfnamefont {C.}~\bibnamefont
  {Janot}},\ }\href {https://books.google.es/books?id=dpPePU-xs2oC} {\emph
  {\bibinfo {title} {Quasicrystals: A Primer}}},\ Monographs on the physics and
  chemistry of materials\ (\bibinfo  {publisher} {Clarendon Press},\ \bibinfo
  {year} {1997})\BibitemShut {NoStop}%
\bibitem [{\citenamefont {Janssen}\ \emph {et~al.}(2007)\citenamefont
  {Janssen}, \citenamefont {Chapuis}, \citenamefont {De~Boissieu} \emph
  {et~al.}}]{janssen2007aperiodic}%
  \BibitemOpen
  \bibfield  {author} {\bibinfo {author} {\bibfnamefont {T.}~\bibnamefont
  {Janssen}}, \bibinfo {author} {\bibfnamefont {G.}~\bibnamefont {Chapuis}},
  \bibinfo {author} {\bibfnamefont {M.}~\bibnamefont {De~Boissieu}},  \emph
  {et~al.},\ }\href@noop {} {\emph {\bibinfo {title} {Aperiodic crystals: from
  modulated phases to quasicrystals}}},\ Vol.~\bibinfo {volume} {20}\ (\bibinfo
   {publisher} {Oxford University Press},\ \bibinfo {year} {2007})\BibitemShut
  {NoStop}%
\bibitem [{\citenamefont {Stadnik}(2012)}]{stadnik2012physical}%
  \BibitemOpen
  \bibfield  {author} {\bibinfo {author} {\bibfnamefont {Z.}~\bibnamefont
  {Stadnik}},\ }\href {https://books.google.es/books?id=AFrrCAAAQBAJ} {\emph
  {\bibinfo {title} {Physical Properties of Quasicrystals}}},\ Springer Series
  in Solid-State Sciences\ (\bibinfo  {publisher} {Springer Berlin
  Heidelberg},\ \bibinfo {year} {2012})\BibitemShut {NoStop}%
\bibitem [{\citenamefont {Fan}(2016)}]{fan2016mathematical}%
  \BibitemOpen
  \bibfield  {author} {\bibinfo {author} {\bibfnamefont {T.}~\bibnamefont
  {Fan}},\ }\href {https://books.google.es/books?id=V80cDQAAQBAJ} {\emph
  {\bibinfo {title} {Mathematical Theory of Elasticity of Quasicrystals and Its
  Applications}}},\ Springer Series in Materials Science\ (\bibinfo
  {publisher} {Springer Singapore},\ \bibinfo {year} {2016})\BibitemShut
  {NoStop}%
\bibitem [{\citenamefont {Scott}\ and\ \citenamefont
  {Clark}(2012)}]{scott2012incommensurate}%
  \BibitemOpen
  \bibfield  {author} {\bibinfo {author} {\bibfnamefont {J.}~\bibnamefont
  {Scott}}\ and\ \bibinfo {author} {\bibfnamefont {N.}~\bibnamefont {Clark}},\
  }\href {https://books.google.es/books?id=QYngBwAAQBAJ} {\emph {\bibinfo
  {title} {Incommensurate Crystals, Liquid Crystals, and Quasi-Crystals}}},\
  Nato Science Series B:\ (\bibinfo  {publisher} {Springer US},\ \bibinfo
  {year} {2012})\BibitemShut {NoStop}%
\bibitem [{\citenamefont {Jari{\'c}}\ \emph {et~al.}(1988)\citenamefont
  {Jari{\'c}}, \citenamefont {Jaric}, \citenamefont {Bak},\ and\ \citenamefont
  {Gratias}}]{jaric1988introduction}%
  \BibitemOpen
  \bibfield  {author} {\bibinfo {author} {\bibfnamefont {M.}~\bibnamefont
  {Jari{\'c}}}, \bibinfo {author} {\bibfnamefont {M.}~\bibnamefont {Jaric}},
  \bibinfo {author} {\bibfnamefont {P.}~\bibnamefont {Bak}}, \ and\ \bibinfo
  {author} {\bibfnamefont {D.}~\bibnamefont {Gratias}},\ }\href
  {https://www.elsevier.com/books/introduction-to-quasicrystals/jaric/978-0-12-040601-2}
  {\emph {\bibinfo {title} {Introduction to Quasicrystals}}}\ (\bibinfo
  {publisher} {Academic Press},\ \bibinfo {year} {1988})\BibitemShut {NoStop}%
\bibitem [{\citenamefont {Janssen}\ and\ \citenamefont
  {Janner}(2014)}]{janssen2014aperiodic}%
  \BibitemOpen
  \bibfield  {author} {\bibinfo {author} {\bibfnamefont {T.}~\bibnamefont
  {Janssen}}\ and\ \bibinfo {author} {\bibfnamefont {A.}~\bibnamefont
  {Janner}},\ }\href@noop {} {\bibfield  {journal} {\bibinfo  {journal} {Acta
  Crystallographica Section B: Structural Science, Crystal Engineering and
  Materials}\ }\textbf {\bibinfo {volume} {70}},\ \bibinfo {pages} {617}
  (\bibinfo {year} {2014})}\BibitemShut {NoStop}%
\bibitem [{\citenamefont {de~Boissieu}(2012)}]{C2CS35212E}%
  \BibitemOpen
  \bibfield  {author} {\bibinfo {author} {\bibfnamefont {M.}~\bibnamefont
  {de~Boissieu}},\ }\href {\doibase 10.1039/C2CS35212E} {\bibfield  {journal}
  {\bibinfo  {journal} {Chem. Soc. Rev.}\ }\textbf {\bibinfo {volume} {41}},\
  \bibinfo {pages} {6778} (\bibinfo {year} {2012})}\BibitemShut {NoStop}%
\bibitem [{\citenamefont {Janssen}\ \emph {et~al.}(2002)\citenamefont
  {Janssen}, \citenamefont {Radulescu},\ and\ \citenamefont
  {Rubtsov}}]{Janssen2002}%
  \BibitemOpen
  \bibfield  {author} {\bibinfo {author} {\bibfnamefont {T.}~\bibnamefont
  {Janssen}}, \bibinfo {author} {\bibfnamefont {O.}~\bibnamefont {Radulescu}},
  \ and\ \bibinfo {author} {\bibfnamefont {A.~N.}\ \bibnamefont {Rubtsov}},\
  }\href {\doibase 10.1140/epjb/e2002-00265-y} {\bibfield  {journal} {\bibinfo
  {journal} {The European Physical Journal B - Condensed Matter and Complex
  Systems}\ }\textbf {\bibinfo {volume} {29}},\ \bibinfo {pages} {85} (\bibinfo
  {year} {2002})}\BibitemShut {NoStop}%
\bibitem [{\citenamefont {Socolar}\ \emph {et~al.}(1986)\citenamefont
  {Socolar}, \citenamefont {Lubensky},\ and\ \citenamefont
  {Steinhardt}}]{PhysRevB.34.3345}%
  \BibitemOpen
  \bibfield  {author} {\bibinfo {author} {\bibfnamefont {J.~E.~S.}\
  \bibnamefont {Socolar}}, \bibinfo {author} {\bibfnamefont {T.~C.}\
  \bibnamefont {Lubensky}}, \ and\ \bibinfo {author} {\bibfnamefont {P.~J.}\
  \bibnamefont {Steinhardt}},\ }\href {\doibase 10.1103/PhysRevB.34.3345}
  {\bibfield  {journal} {\bibinfo  {journal} {Phys. Rev. B}\ }\textbf {\bibinfo
  {volume} {34}},\ \bibinfo {pages} {3345} (\bibinfo {year}
  {1986})}\BibitemShut {NoStop}%
\bibitem [{\citenamefont {[de Boissieu]}\ \emph {et~al.}(2008)\citenamefont
  {[de Boissieu]}, \citenamefont {Currat},\ and\ \citenamefont
  {Francoual}}]{DEBOISSIEU2008107}%
  \BibitemOpen
  \bibfield  {author} {\bibinfo {author} {\bibfnamefont {M.}~\bibnamefont {[de
  Boissieu]}}, \bibinfo {author} {\bibfnamefont {R.}~\bibnamefont {Currat}}, \
  and\ \bibinfo {author} {\bibfnamefont {S.}~\bibnamefont {Francoual}},\ }in\
  \href {\doibase https://doi.org/10.1016/S1570-002X(08)80020-1} {\emph
  {\bibinfo {booktitle} {Quasicrystals}}},\ \bibinfo {series} {Handbook of
  Metal Physics}, Vol.~\bibinfo {volume} {3},\ \bibinfo {editor} {edited by\
  \bibinfo {editor} {\bibfnamefont {T.}~\bibnamefont {Fujiwara}}\ and\ \bibinfo
  {editor} {\bibfnamefont {Y.}~\bibnamefont {Ishii}}}\ (\bibinfo  {publisher}
  {Elsevier},\ \bibinfo {year} {2008})\ pp.\ \bibinfo {pages} {107 --
  169}\BibitemShut {NoStop}%
\bibitem [{\citenamefont {Bak}(1985)}]{PhysRevLett.54.1517}%
  \BibitemOpen
  \bibfield  {author} {\bibinfo {author} {\bibfnamefont {P.}~\bibnamefont
  {Bak}},\ }\href {\doibase 10.1103/PhysRevLett.54.1517} {\bibfield  {journal}
  {\bibinfo  {journal} {Phys. Rev. Lett.}\ }\textbf {\bibinfo {volume} {54}},\
  \bibinfo {pages} {1517} (\bibinfo {year} {1985})}\BibitemShut {NoStop}%
\bibitem [{\citenamefont {Widom}(2008)}]{widom2008discussion}%
  \BibitemOpen
  \bibfield  {author} {\bibinfo {author} {\bibfnamefont {M.}~\bibnamefont
  {Widom}},\ }\href@noop {} {\bibfield  {journal} {\bibinfo  {journal}
  {Philosophical Magazine}\ }\textbf {\bibinfo {volume} {88}},\ \bibinfo
  {pages} {2339} (\bibinfo {year} {2008})}\BibitemShut {NoStop}%
\bibitem [{\citenamefont {Zeyher}\ and\ \citenamefont
  {Finger}(1982)}]{PhysRevLett.49.1833}%
  \BibitemOpen
  \bibfield  {author} {\bibinfo {author} {\bibfnamefont {R.}~\bibnamefont
  {Zeyher}}\ and\ \bibinfo {author} {\bibfnamefont {W.}~\bibnamefont
  {Finger}},\ }\href {\doibase 10.1103/PhysRevLett.49.1833} {\bibfield
  {journal} {\bibinfo  {journal} {Phys. Rev. Lett.}\ }\textbf {\bibinfo
  {volume} {49}},\ \bibinfo {pages} {1833} (\bibinfo {year}
  {1982})}\BibitemShut {NoStop}%
\bibitem [{\citenamefont {Pekker}\ and\ \citenamefont
  {Varma}(2015)}]{doi:10.1146/annurev-conmatphys-031214-014350}%
  \BibitemOpen
  \bibfield  {author} {\bibinfo {author} {\bibfnamefont {D.}~\bibnamefont
  {Pekker}}\ and\ \bibinfo {author} {\bibfnamefont {C.}~\bibnamefont {Varma}},\
  }\href {\doibase 10.1146/annurev-conmatphys-031214-014350} {\bibfield
  {journal} {\bibinfo  {journal} {Annual Review of Condensed Matter Physics}\
  }\textbf {\bibinfo {volume} {6}},\ \bibinfo {pages} {269} (\bibinfo {year}
  {2015})},\ \Eprint
  {http://arxiv.org/abs/https://doi.org/10.1146/annurev-conmatphys-031214-014350}
  {https://doi.org/10.1146/annurev-conmatphys-031214-014350} \BibitemShut
  {NoStop}%
\bibitem [{\citenamefont {Francoual}\ \emph {et~al.}(2003)\citenamefont
  {Francoual}, \citenamefont {Livet}, \citenamefont {de~Boissieu},
  \citenamefont {Yakhou}, \citenamefont {Bley}, \citenamefont {L\'etoublon},
  \citenamefont {Caudron},\ and\ \citenamefont
  {Gastaldi}}]{PhysRevLett.91.225501}%
  \BibitemOpen
  \bibfield  {author} {\bibinfo {author} {\bibfnamefont {S.}~\bibnamefont
  {Francoual}}, \bibinfo {author} {\bibfnamefont {F.}~\bibnamefont {Livet}},
  \bibinfo {author} {\bibfnamefont {M.}~\bibnamefont {de~Boissieu}}, \bibinfo
  {author} {\bibfnamefont {F.}~\bibnamefont {Yakhou}}, \bibinfo {author}
  {\bibfnamefont {F.}~\bibnamefont {Bley}}, \bibinfo {author} {\bibfnamefont
  {A.}~\bibnamefont {L\'etoublon}}, \bibinfo {author} {\bibfnamefont
  {R.}~\bibnamefont {Caudron}}, \ and\ \bibinfo {author} {\bibfnamefont
  {J.}~\bibnamefont {Gastaldi}},\ }\href {\doibase
  10.1103/PhysRevLett.91.225501} {\bibfield  {journal} {\bibinfo  {journal}
  {Phys. Rev. Lett.}\ }\textbf {\bibinfo {volume} {91}},\ \bibinfo {pages}
  {225501} (\bibinfo {year} {2003})}\BibitemShut {NoStop}%
\bibitem [{\citenamefont {Durand}\ \emph {et~al.}(1991)\citenamefont {Durand},
  \citenamefont {Papoular}, \citenamefont {Currat}, \citenamefont {Lambert},
  \citenamefont {Legrand},\ and\ \citenamefont
  {Mezei}}]{durand1991investigation}%
  \BibitemOpen
  \bibfield  {author} {\bibinfo {author} {\bibfnamefont {D.}~\bibnamefont
  {Durand}}, \bibinfo {author} {\bibfnamefont {R.}~\bibnamefont {Papoular}},
  \bibinfo {author} {\bibfnamefont {R.}~\bibnamefont {Currat}}, \bibinfo
  {author} {\bibfnamefont {M.}~\bibnamefont {Lambert}}, \bibinfo {author}
  {\bibfnamefont {J.}~\bibnamefont {Legrand}}, \ and\ \bibinfo {author}
  {\bibfnamefont {F.}~\bibnamefont {Mezei}},\ }\href@noop {} {\bibfield
  {journal} {\bibinfo  {journal} {Physical Review B}\ }\textbf {\bibinfo
  {volume} {43}},\ \bibinfo {pages} {10690} (\bibinfo {year}
  {1991})}\BibitemShut {NoStop}%
\bibitem [{\citenamefont {Lubensky}\ \emph {et~al.}(1985)\citenamefont
  {Lubensky}, \citenamefont {Ramaswamy},\ and\ \citenamefont
  {Toner}}]{PhysRevB.32.7444}%
  \BibitemOpen
  \bibfield  {author} {\bibinfo {author} {\bibfnamefont {T.~C.}\ \bibnamefont
  {Lubensky}}, \bibinfo {author} {\bibfnamefont {S.}~\bibnamefont {Ramaswamy}},
  \ and\ \bibinfo {author} {\bibfnamefont {J.}~\bibnamefont {Toner}},\ }\href
  {\doibase 10.1103/PhysRevB.32.7444} {\bibfield  {journal} {\bibinfo
  {journal} {Phys. Rev. B}\ }\textbf {\bibinfo {volume} {32}},\ \bibinfo
  {pages} {7444} (\bibinfo {year} {1985})}\BibitemShut {NoStop}%
\bibitem [{\citenamefont {Baggioli}\ \emph {et~al.}(2020)\citenamefont
  {Baggioli}, \citenamefont {Brazhkin}, \citenamefont {Trachenko},\ and\
  \citenamefont {Vasin}}]{Baggioli:2019jcm}%
  \BibitemOpen
  \bibfield  {author} {\bibinfo {author} {\bibfnamefont {M.}~\bibnamefont
  {Baggioli}}, \bibinfo {author} {\bibfnamefont {V.~V.}\ \bibnamefont
  {Brazhkin}}, \bibinfo {author} {\bibfnamefont {K.}~\bibnamefont {Trachenko}},
  \ and\ \bibinfo {author} {\bibfnamefont {M.}~\bibnamefont {Vasin}},\ }\href
  {\doibase 10.1016/j.physrep.2020.04.002} {\bibfield  {journal} {\bibinfo
  {journal} {Phys. Rept.}\ }\textbf {\bibinfo {volume} {865}},\ \bibinfo
  {pages} {1} (\bibinfo {year} {2020})},\ \Eprint
  {http://arxiv.org/abs/1904.01419} {arXiv:1904.01419 [cond-mat.stat-mech]}
  \BibitemShut {NoStop}%
\bibitem [{\citenamefont {Baggioli}\ and\ \citenamefont {Landry}()}]{toappear}%
  \BibitemOpen
  \bibfield  {author} {\bibinfo {author} {\bibfnamefont {M.}~\bibnamefont
  {Baggioli}}\ and\ \bibinfo {author} {\bibfnamefont {M.}~\bibnamefont
  {Landry}},\ }\href@noop {} {\bibinfo  {journal} {To appear.}\ }\BibitemShut
  {NoStop}%
\bibitem [{\citenamefont {Chaikin}\ and\ \citenamefont
  {Lubensky}(2000)}]{chaikin2000principles}%
  \BibitemOpen
\bibfield  {journal} {  }\bibfield  {author} {\bibinfo {author} {\bibfnamefont
  {P.}~\bibnamefont {Chaikin}}\ and\ \bibinfo {author} {\bibfnamefont
  {T.}~\bibnamefont {Lubensky}},\ }\href
  {https://books.google.gr/books?id=P9YjNjzr9OIC} {\emph {\bibinfo {title}
  {Principles of Condensed Matter Physics}}}\ (\bibinfo  {publisher} {Cambridge
  University Press},\ \bibinfo {year} {2000})\BibitemShut {NoStop}%
\bibitem [{\citenamefont {Nika}\ and\ \citenamefont
  {Balandin}(2017)}]{Nika_2017}%
  \BibitemOpen
  \bibfield  {author} {\bibinfo {author} {\bibfnamefont {D.~L.}\ \bibnamefont
  {Nika}}\ and\ \bibinfo {author} {\bibfnamefont {A.~A.}\ \bibnamefont
  {Balandin}},\ }\href {\doibase 10.1088/1361-6633/80/3/036502} {\bibfield
  {journal} {\bibinfo  {journal} {Reports on Progress in Physics}\ }\textbf
  {\bibinfo {volume} {80}},\ \bibinfo {pages} {036502} (\bibinfo {year}
  {2017})}\BibitemShut {NoStop}%
\bibitem [{\citenamefont {Esposito}\ \emph {et~al.}(2020)\citenamefont
  {Esposito}, \citenamefont {Geoffray},\ and\ \citenamefont
  {Melia}}]{Esposito:2020hwq}%
  \BibitemOpen
  \bibfield  {author} {\bibinfo {author} {\bibfnamefont {A.}~\bibnamefont
  {Esposito}}, \bibinfo {author} {\bibfnamefont {E.}~\bibnamefont {Geoffray}},
  \ and\ \bibinfo {author} {\bibfnamefont {T.}~\bibnamefont {Melia}},\
  }\href@noop {} {\  (\bibinfo {year} {2020})},\ \Eprint
  {http://arxiv.org/abs/2006.05429} {arXiv:2006.05429 [hep-th]} \BibitemShut
  {NoStop}%
\bibitem [{\citenamefont {Shimano}\ and\ \citenamefont
  {Tsuji}(2020)}]{doi:10.1146/annurev-conmatphys-031119-050813}%
  \BibitemOpen
  \bibfield  {author} {\bibinfo {author} {\bibfnamefont {R.}~\bibnamefont
  {Shimano}}\ and\ \bibinfo {author} {\bibfnamefont {N.}~\bibnamefont
  {Tsuji}},\ }\href {\doibase 10.1146/annurev-conmatphys-031119-050813}
  {\bibfield  {journal} {\bibinfo  {journal} {Annual Review of Condensed Matter
  Physics}\ }\textbf {\bibinfo {volume} {11}},\ \bibinfo {pages} {103}
  (\bibinfo {year} {2020})},\ \Eprint
  {http://arxiv.org/abs/https://doi.org/10.1146/annurev-conmatphys-031119-050813}
  {https://doi.org/10.1146/annurev-conmatphys-031119-050813} \BibitemShut
  {NoStop}%
\bibitem [{\citenamefont {Allen}\ \emph {et~al.}(1999)\citenamefont {Allen},
  \citenamefont {Feldman}, \citenamefont {Fabian},\ and\ \citenamefont
  {Wooten}}]{allen1999diffusons}%
  \BibitemOpen
  \bibfield  {author} {\bibinfo {author} {\bibfnamefont {P.~B.}\ \bibnamefont
  {Allen}}, \bibinfo {author} {\bibfnamefont {J.~L.}\ \bibnamefont {Feldman}},
  \bibinfo {author} {\bibfnamefont {J.}~\bibnamefont {Fabian}}, \ and\ \bibinfo
  {author} {\bibfnamefont {F.}~\bibnamefont {Wooten}},\ }\href@noop {}
  {\bibfield  {journal} {\bibinfo  {journal} {Philosophical Magazine B}\
  }\textbf {\bibinfo {volume} {79}},\ \bibinfo {pages} {1715} (\bibinfo {year}
  {1999})}\BibitemShut {NoStop}%
\bibitem [{\citenamefont {Moratalla}\ \emph
  {et~al.}(2019{\natexlab{b}})\citenamefont {Moratalla}, \citenamefont
  {Gebbia}, \citenamefont {Ramos}, \citenamefont {Pardo}, \citenamefont
  {Mukhopadhyay}, \citenamefont {Rudi\ifmmode~\acute{c}\else \'{c}\fi{}},
  \citenamefont {Fernandez-Alonso}, \citenamefont {Bermejo},\ and\
  \citenamefont {Tamarit}}]{Moratalla}%
  \BibitemOpen
  \bibfield  {author} {\bibinfo {author} {\bibfnamefont {M.}~\bibnamefont
  {Moratalla}}, \bibinfo {author} {\bibfnamefont {J.~F.}\ \bibnamefont
  {Gebbia}}, \bibinfo {author} {\bibfnamefont {M.~A.}\ \bibnamefont {Ramos}},
  \bibinfo {author} {\bibfnamefont {L.~C.}\ \bibnamefont {Pardo}}, \bibinfo
  {author} {\bibfnamefont {S.}~\bibnamefont {Mukhopadhyay}}, \bibinfo {author}
  {\bibfnamefont {S.}~\bibnamefont {Rudi\ifmmode~\acute{c}\else \'{c}\fi{}}},
  \bibinfo {author} {\bibfnamefont {F.}~\bibnamefont {Fernandez-Alonso}},
  \bibinfo {author} {\bibfnamefont {F.~J.}\ \bibnamefont {Bermejo}}, \ and\
  \bibinfo {author} {\bibfnamefont {J.~L.}\ \bibnamefont {Tamarit}},\ }\href
  {\doibase 10.1103/PhysRevB.99.024301} {\bibfield  {journal} {\bibinfo
  {journal} {Phys. Rev. B}\ }\textbf {\bibinfo {volume} {99}},\ \bibinfo
  {pages} {024301} (\bibinfo {year} {2019}{\natexlab{b}})}\BibitemShut
  {NoStop}%
\bibitem [{\citenamefont {Chernikov}\ \emph {et~al.}(1995)\citenamefont
  {Chernikov}, \citenamefont {Bianchi},\ and\ \citenamefont
  {Ott}}]{PhysRevB.51.153}%
  \BibitemOpen
  \bibfield  {author} {\bibinfo {author} {\bibfnamefont {M.~A.}\ \bibnamefont
  {Chernikov}}, \bibinfo {author} {\bibfnamefont {A.}~\bibnamefont {Bianchi}},
  \ and\ \bibinfo {author} {\bibfnamefont {H.~R.}\ \bibnamefont {Ott}},\ }\href
  {\doibase 10.1103/PhysRevB.51.153} {\bibfield  {journal} {\bibinfo  {journal}
  {Phys. Rev. B}\ }\textbf {\bibinfo {volume} {51}},\ \bibinfo {pages} {153}
  (\bibinfo {year} {1995})}\BibitemShut {NoStop}%
\bibitem [{\citenamefont {Popcevic}\ \emph {et~al.}(2011)\citenamefont
  {Popcevic}, \citenamefont {Stanic}, \citenamefont {Bihar}, \citenamefont
  {Bilusic},\ and\ \citenamefont {Smontara}}]{doi:10.1002/ijch.201100150}%
  \BibitemOpen
  \bibfield  {author} {\bibinfo {author} {\bibfnamefont {P.}~\bibnamefont
  {Popcevic}}, \bibinfo {author} {\bibfnamefont {D.}~\bibnamefont {Stanic}},
  \bibinfo {author} {\bibfnamefont {Z.}~\bibnamefont {Bihar}}, \bibinfo
  {author} {\bibfnamefont {A.}~\bibnamefont {Bilusic}}, \ and\ \bibinfo
  {author} {\bibfnamefont {A.}~\bibnamefont {Smontara}},\ }\href {\doibase
  10.1002/ijch.201100150} {\bibfield  {journal} {\bibinfo  {journal} {Israel
  Journal of Chemistry}\ }\textbf {\bibinfo {volume} {51}},\ \bibinfo {pages}
  {1340} (\bibinfo {year} {2011})},\ \Eprint
  {http://arxiv.org/abs/https://onlinelibrary.wiley.com/doi/pdf/10.1002/ijch.201100150}
  {https://onlinelibrary.wiley.com/doi/pdf/10.1002/ijch.201100150} \BibitemShut
  {NoStop}%
\bibitem [{\citenamefont {Ochoa}(2019)}]{PhysRevB.100.155426}%
  \BibitemOpen
  \bibfield  {author} {\bibinfo {author} {\bibfnamefont {H.}~\bibnamefont
  {Ochoa}},\ }\href {\doibase 10.1103/PhysRevB.100.155426} {\bibfield
  {journal} {\bibinfo  {journal} {Phys. Rev. B}\ }\textbf {\bibinfo {volume}
  {100}},\ \bibinfo {pages} {155426} (\bibinfo {year} {2019})}\BibitemShut
  {NoStop}%
\end{thebibliography}%

\appendix
\end{document}